\documentclass[bibyear]{aa}  
\usepackage[colorlinks=true,allcolors=blue]{hyperref}
\usepackage{natbib}
\usepackage{twoopt}
\bibpunct{(}{)}{;}{a}{}{,} 

\usepackage{graphicx}
\usepackage{linenoaa}
\usepackage{txfonts}

\makeatletter
\renewcommand*\aa@pageof{, page \thepage{} of \pageref*{LastPage}}
\makeatother

\usepackage{soul}

\begin{document} 

\title{Dark energy and negative compressibility of the Universe: A study based on cosmic equations of state}
\titlerunning{Cosmological equations of state and its compressibility properties }

 \author{Felipe A. Asenjo
          \inst{1}
          \and
          \'Osmar Rodr\'iguez
          \inst{2}
          \and
         Ariel \'Ordenes Morales
          \inst{2}
          \and
         Alejandro Clocchiatti
          \inst{2}
          }
   \institute{Facultad de Ingenier\'{\i}a y Ciencias, Universidad Adolfo Ib\'a\~nez, Santiago 7941169, Chile\\
              \email{felipe.asenjo@uai.cl}
         \and
           Pontificia Universidad Cat\'olica de Chile, Vicu\~na Mackenna 4860, Macul, Santiago, Chile
            }
    
   \date{Received ; accepted}
  
  \abstract
   {All cosmological models are defined through energy and matter, which describe the evolution of the targeted Universe. Whether the content is simple or complex, it can be defined through an effective equation of state (EoS).}
  {We study different cosmological models by analyzing the properties of the complete EoS of their fluid contents.}   
   {We obtained general expressions for the ratio between pressure $p$ and energy density $\rho$, and for their derivative $dp/d\rho$ in terms of deceleration and jerk parameters. We show that, in general, a cosmology whose energy density is constructed from multiple ideal gases cannot be described by the EoS of a single, ideal gas. This is directly related to a variation in the deceleration parameter during the evolution of the Universe. Furthermore, we studied the effective compressibility properties of the EoS, demonstrating that dark energy features of the fluid should be related to the negative compressibility of the contents. These features depend on the dynamics of the jerk parameter for each cosmological system, implying that the knowledge of dark energy evolution is encoded in the third-order derivatives of the scale factor.}
   {This approach to understanding the properties of cosmological models indicates that the accelerating evolution of some models, such as the  lambda-cold dark matter model ($\Lambda$CDM) or the Chaplygin gas model, is due to the EoS of their nonideal content fluids rather than an associated anomalous (dark energy) nature. On the contrary, models such as  $w$-cold dark matter ($w$CDM), $w_0$$w_a$-cold dark matter ($w_0$$w_a$CDM), or Dvali-Gabadadze-Porrati models, include by construction fluids with anomalous content, implying a negative compressibility and an effective dark energy behavior. }
   {Our work highlights the importance of treating all contents of the Universe as a nonideal fluid system, with a unique EoS, from which properties such as dark energy can be inferred.}


\maketitle

\nolinenumbers 

\section{Introduction}

In general, a fluid with density $n$ (particles per unit volume), pressure $p$, and
energy density $\rho$ 
can be characterized by an equation of state (EoS), for instance, in the form $p=p(n)$ or $p=p(\rho)$. In such a case, different properties of such a fluid can be studied from the EoS. One of these properties is fluid compressibility,
\begin{equation}
\beta=\frac{1}{n}\frac{d n}{dp},
\end{equation}
which quantifies how fluid volume changes in response to a change in pressure \citep{zemansky}.
Ordinary, natural fluids have positive compressibility $\beta>0$ (since $n$ is positive). This means, straightforwardly, that an increase in energy density implies an increase in pressure and vice versa.
This behavior is expected in ordinary fluids as it is a condition for their mechanical stability.

However, certain anomalous materials and fluids exhibit $\beta<0$, i.e., a behavior that contrasts with the usual notion of compressibility. This property, called negative compressibility \citep{Miller,Gatt,Cairns,Schakel}, implies that anomalous fluids experience an increase in pressure with decreasing energy density, or vice versa.
In other words, the fluid volume increases as pressure increases.  
These anomalous materials are studied in the realm of metamaterials \citep{zachary2,Ime,Zachary,Xie,Caprini,Baughman,Arash,grima2,qu,Lu,weili} in the search for materials with novel properties.

In cosmology, the content of the Universe is typically modeled through its energy density and pressure.
Given the EoS of the Universe's content, its compressibility properties can be obtained through its connection with the square of the effective adiabatic sound speed, defined as the change in pressure with respect to energy density, 
${dp}/{d\rho}$. Hence, 
we can write their relation as
\begin{equation}
\frac{d\rho}{dp}=\beta\, n \frac{d\rho}{dn}\, .
\end{equation}
Since for all fluids $n(d\rho/dn)>0$, any ordinary fluid with positive compressibility $\beta>0$ has  $d\rho/dp>0$. By contrast, 
anomalous fluids with negative compressibility, $\beta<0$, have $d\rho/dp<0$.

In this view, the typical ideal fluids used in cosmology to model the contents of the universe have EoSs of the form 
\begin{equation}
p=w\rho,
\label{eq:ideal_fluid}
\end{equation}
with constant $w$. Therefore, ordinary ideal fluids with $w>0$ have positive compressibility behaviors, $dp/d\rho=w>0$.
On the other hand, ideal dark energy fluids are characterized by $w<0$, where in particular, the cosmological constant corresponds to $w=-1$.
Hence, an ideal dark energy fluid has $dp/d\rho=w<0$, even when the energy density and pressure do not depend on time.  
Notice that negative compressibility properties related to dark energies are not due to negative pressure, since this latter value could result from specific shifts in scale.
Thus, unusual dark energy properties can arise directly from negative compressibility properties, implying that the fluid exerts greater pressure as its energy density decreases.

Therefore, cosmological models can be studied through the EoS of their contents by analyzing compressibility properties.
This is especially relevant when dark energy is proposed as a consequence of fitting models, such as Eq.~\eqref{eq:ideal_fluid}, to the observed data. It is even more important when the Universe is assumed to be composed of multiple fluids, as is the case in current standard cosmological models.

 In that spirit, we show in Sect.~\ref{se:general_fluids} that it  is always possible to calculate the $dp/d\rho$ of the fluid content for any cosmological model, and, thus, to relate it to the compressibility properties of the model.
In Sect.~\ref{modelosteoricos}, we study specific cosmological systems to show how they define a unique EoS of the entire fluid content of their universes. We then use the compressibility properties of these EoSs to show how their attributes differ from those of the ideal fluids from which they were originally constructed. We also relate their compressibility properties with different dark energy proposals, showing how their standard definitions differ from the results obtained when assuming compressibility.
Finally, in Sect.~\ref{se:discussion}, we present a brief discussion of our findings, emphasizing the importance of different cosmological parameters (Hubble, acceleration, and jerk) in characterizing ordinary or anomalous fluid properties.
 
\section{General fluid compressibility in cosmology}
\label{se:general_fluids}

When the Universe is modeled as an isotropic homogeneous entity, its cosmological evolution can be obtained from
Einstein equations using the Friedmann-Lemaître-Robertson-Walker (FLRW) metric.  The evolution of this metric can be reduced to the two following equations \citep{weinberg}:
\begin{eqnarray}
    H^2&=&\frac{8\pi G}{3}\rho\, , \label{eq1}\\
    \dot\rho&=&-3H(\rho+p)\, ,\label{eq2}
\end{eqnarray}
where $H=\dot a/a$ is the Hubble parameter, the dot ($\, \dot{}\,$) denotes the derivative with respect to time, $a=a(t)$ is the scale factor (which increases with time for an expanding Universe), and $G$ is the gravitational constant. We adopt units, such that the speed of light $c=1$. Finally, from Eqs.~\eqref{eq1} and \eqref{eq2}, a third dependent equation can obtained:
\begin{equation}
    \frac{\ddot a}{a}=-\frac{4\pi G}{3}(\rho+3p)\, .
    \label{eqaddq}
\end{equation}
This method is a standard way to model the homogeneous content of the Universe as perfect fluid, with energy densities $\rho=\rho(a)=\rho(t)$ and pressure $p=p(a)=p(t)$. Equation~\eqref{eq1} shows how the Universe expands as a result of its content. Equation~\eqref{eq2} establishes that the energy content is conserved during the evolution of the Universe. These two equations serve as the starting point for all the preferred cosmological models 
used today.

It is important to note that Eqs.~\eqref{eq1} and \eqref{eq2} are written
 under the sole assumption that the Universe has a unique energy density, $\rho$, and pressure, $p$. Therefore, any interpretation of the Universe's content must follow that premise.
Currently, the most widely used approach in cosmology involves proposing a fitting model for  Eq.~\eqref{eq1} in terms of an ansatz for the energy density as a function of the scale factor.
On the contrary,  the continuity equation, Eq. \eqref{eq2}, is rarely studied to understand the temporal dynamics of pressure.
Even less studied is the EoS of the fluid content derived from such an ansatz.

For any given fitting model of the energy density expressed as a function of the scale factor, the fluid pressure can be obtained from Eq.~\eqref{eq2} via
\begin{equation}
    p=-\frac{a}{3}\frac{d\rho}{da}-\rho\, .
    \label{equooressureGenrrr}
\end{equation}
Therefore, when the functional form $\rho=\rho(a)$ of the energy density is known, the corresponding functional form $p=p(a)$ can be obtained. In principle, Eq.~\eqref{eq1} allow us to obtain $a=a(\rho)$. Therefore, we can use
Eq.~\eqref{equooressureGenrrr} to obtain the general EoS, $p=p(\rho)$, of any cosmological system.

It is important to note that Eq.~\eqref{equooressureGenrrr} is equivalent to the first law of thermodynamics, $p=n(\partial\rho/\partial n)-\rho$, for a perfect fluid in curved space-time, where density $n\propto a^{-3}$ \citep{misner}. Therefore, Eq.~\eqref{equooressureGenrrr}  establishes that any perfect fluid used as   the content of a Universe described by the FLRW Eqs.~\eqref{eq1} and \eqref{eq2}, behaves as an adiabatic flow.

Although Eq.~\eqref{equooressureGenrrr}  allows us to deduce the EoS of the fluid, in cosmology, the ratio $p/\rho$ is typically calculated instead. We 
can use Eqs.~\eqref{eq1} and \eqref{eq2} to establish that $\dot H=-4\pi G (\rho+p)$. This allows us to obtain the relation
\begin{eqnarray}
    \frac{p}{\rho}=\frac{2q-1}{3}\, ,
    \label{relationprhodivi}
\end{eqnarray}
where $q=-a\ddot a/\dot a^2$ is the deceleration parameter. Equation~\eqref{relationprhodivi} is not an EoS, as it does not provide the functional form $p=p(\rho)$, unless we are dealing with a single, ideal gas.
It tells us that $p/\rho$ is not constant in time and changes as $q$ evolves. Therefore, any perfect fluid used in Einstein equations, Eqs. \eqref{eq1} and \eqref{eq2} is a  fluid with an EoS that may differ
from the ideal gas approximation, where $p/\rho\neq\mbox{constant}$. This is relevant to any model used to fit cosmological data, especially when the model consists of the sum of several ideal gases invoking Dalton's law of thermodynamics \citep{Missen}.

Furthermore, $p/\rho>0$ only for cosmological models with $q>1/2$. In the case that $q<1/2$, such a ratio becomes negative, and it defines a specific cosmological time at which the pressure becomes negative with respect to the energy density. It neither provides any additional information about the fluid's behavior in terms of the EoS nor if the pressure increases or decreases.

On the other hand, the sign of fluid compressibility can be readily obtained by calculating $dp/d\rho=(dp/da)(d\rho/da)^{-1}$. This can be generally expressed as
\begin{equation}
    \frac{dp}{d\rho}=-\frac{4}{3}-\frac{a}{3}\left(\frac{d\rho}{da} \right)^{-1}\left( \frac{d^2\rho}{da^2}\right)\, .
    \label{GeneralgeneralEOS}
\end{equation}
This expression can be explicitly evaluated using Eq.~\eqref{eq1}. We first obtain
\begin{equation}
    \frac{d\rho}{da}=-\frac{3 H^2}{4\pi G\, a}(1+q)\, .
\end{equation}
Similarly, we find
\begin{equation}
    \frac{d^2\rho}{da^2}=\frac{3 H^2}{4\pi G\, a^2}(3+4q+j)\, ,
\end{equation}
where $j=a^2\, \dddot a/\dot a^3$ is the jerk parameter. Hence, the sign of the compressibility \eqref{GeneralgeneralEOS} results from the simple evolving rate
\begin{eqnarray}
    \frac{dp}{d\rho}=\frac{j-1}{3(1+q)}\, .
    \label{GeneralgeneralEOS2}
\end{eqnarray}
The change in pressure with respect to energy density \eqref{GeneralgeneralEOS2} is general and valid for any cosmological model.
We note that 
Eq.~\eqref{GeneralgeneralEOS2} was previously obtained under the assumption of an approximated cosmological EoS, by performing Taylor expansion around the current epoch \citep{visser}.
We stress, however, that the result derived from Eq.~\eqref{GeneralgeneralEOS2} is indeed general and valid at any time.

Different cosmological models yield specific values of $q$ and $j$ (or $H$), depending on how they define the energy density of the Universe, and thereby produce different types of compressibilities. 
Regarding the interpretation that dark energy properties arise from the anomalous behavior of the content EoS, negative compressibility can exist depending not only on the deceleration parameter $q$ but also on the behavior of the jerk parameter $j$.
Dark energy fluids, behaving as anomalous fluids, exist only when $q>-1$ and $j<1$, or $q<-1$ and $j>1$. Otherwise, the fluid content of the Universe behaves normally.
Therefore, a striking conclusion of Eq.
\eqref{GeneralgeneralEOS2} is that, in terms of fluid compressibility properties, the deceleration parameter $q$ alone is insufficient to describe the cosmological behavior of dark energy in any model.
To fully determine the behavior of the fluid content of the Universe -- and, thus, its cosmological evolution -- the jerk parameter must be considered. 
Moreover, surprisingly, there is no need to consider parameters with higher-order derivatives of the scale factor.
Any conclusion drawn from any model based solely on the information given by $q$ will be biased. Thus, the information provided by the jerk parameter is essential.

In terms of a unique, single EoS, an abnormal behavior -- where pressure increases due to a decrease in energy density -- is solely determined by $dp/d\rho<0$.
The abnormal behavior of negative compressibility, as observed for dark energy in an expanding Universe, does not result from negative pressure. Rather, it is the derivative of the EoS, $p=p(\rho)$, which determines the normal or abnormal behavior of a cosmological Universe.

Let us emphasize this in more explicit terms. When a single-component cosmology is proposed -- assuming a single, ideal fluid EoS $p=w\rho$ -- the deceleration parameter is $q=(1+3w)/2$, consistent with Eq.~\eqref{relationprhodivi}. Any accelerated behavior of the scale factor ($\ddot a>0$ or $q<0$) occurs when $w<-1/3$. Therefore, accelerating dark energy satisfies $p/\rho<0$. 
This is true for a single, ideal fluid, as in such cases all the previous definitions contain the same information. However, this is not necessarily valid when the system is described as a multi-fluid. 
By considering Eq.~\eqref{eqaddq}, a general accelerated behavior of the scale factor is only required when 
\begin{equation}
    \rho+3p<0\, .
    \label{rhomas3p}
\end{equation}
This condition (equivalent to $q<0$) generally differs from the negative compressibility condition $dp/d\rho<0$. The former does not measure the changes in energy density and pressure as time evolves. It only indicates when $\ddot a>0$, not its cause. This condition does not provide global information about the evolutionary behavior of the fluid, unlike $dp/d\rho$, which offers detailed insight into the changes in the Universe's content that may later cause acceleration.
For instance, it is possible to construct cosmological models satisfying the condition in Eq. \eqref{rhomas3p}, but with $dp/d \rho>0$ (see the $\Lambda$CDM  model in the following section), such that the Universe's scale factor accelerates even when pressure increases with energy, as in any normal fluid. 
Therefore, as the condition in Eq. \eqref{rhomas3p} ($q<0$ or $\ddot a>0$) can yield different results compared to $dp/d\rho$, it
cannot be a proper definition for dark energy. It is not a general indication of the complete nature of the fluid used in such a model. 

The importance of the effective compressibility effect can be understood by calculating the variance in the deceleration parameter. 
We notice that the FLRW equations, Eqs. \eqref{eq1} and  \eqref{eqaddq}, govern the Hubble and deceleration parameters. They do not establish how the jerk parameter (derivatives of $q$) evolves; therefore, we cannot measure the uncertainty for the evolution of $q$ from them. However, we can calculate the uncertainty using the delta method for the variance in the deceleration parameter, $\mbox{Var}(q)$.  In this case, $\mbox{Var}(q)\approx (\partial q/\partial a)^2\, \mbox{Var}(a)\approx (\dot q/aH)^2\, \mbox{Var}(a)$. Noticing that $\mbox{Var}(H)\approx(\partial H/\partial a)^2 \mbox{Var}(a)\approx H^2(1+q)^2\mbox{Var}(a)/a^2$ and that $\dot q/H=q+2q^2-j=3(1+q)(dp/d\rho-p/\rho)$ (via Eqs.~\eqref{relationprhodivi} and \eqref{GeneralgeneralEOS2}), we find
\begin{equation}
    \mbox{Var}(q)\approx\frac{9}{H^2}\left(\frac{dp}{d\rho}-\frac{p}{\rho}\right)^2\mbox{Var}(H)\, .
    \label{varq}
\end{equation}

Therefore, $dp/d\rho$ is essential for measuring the uncertainties in the models. Of course, this variance cannot be 0, which occurs only in the simplified and unrealistic single-fluid model for an ideal gas (where $dp/d\rho=p/\rho=w$). By contrast, Eq.~\eqref{varq} establishes that multi-fluid or multicomponent models for the Universe have a unique a behavior determined by $dp/d\rho$. Therefore, as $dp/d\rho$ is necessary for calculating the complete behavior of the deceleration parameter, the proper physical nature of the model cannot be contained only in $p/\rho$.

An explicit example is given in Sect.~\ref{modelosteoricos}, where we study several different ansatz frequently used in Eq.~\eqref{eq1}
to model observations of the our expanding Universe within the context of the model-predicted energy density, pressure and, thus, its EoS.
In particular, we analyze the change in pressure with respect to energy density \eqref{GeneralgeneralEOS2} and, thus, the sign of
compressibility in order to characterize the content of the Universe as a normal or an anomalous fluid.  

\section{Approach to different cosmological models}
\label{modelosteoricos}

If a theoretical model is used in Eq.~\eqref{eq1} to fit data, the sign of the compressibility can be analytically determined using Eq.~\eqref{GeneralgeneralEOS2}. In this section we demonstrate this procedure for several known models. We discuss the results for each in the context of the dark energy contribution of the whole fluid.

\subsection{Simple models}

Several fits for cosmological models are constructed under the assumption that the total energy density is the sum of the  contributions from the different epochs of the Universe  with their distinct densities, $\rho=\sum_i\rho_i$.  This is an extension of Dalton's law of thermodynamics for ideal gases \citep{Missen}.
For a single epoch, $i$, filled with a perfect fluid with a barotropic EoS $p_i=w_i\rho_i$ (with a constant value of $-1\leq w_i<1$), the energy density is expressed in terms of the scale factor as $\rho_i=\Omega_i a^{-3(1+w_i)}$, where $\Omega_i>0$ is a constant \citep{weinberg}. This notion may include any form of simple, single dark energy content.
Therefore, the total energy density of the Universe can be modeled as
\begin{equation}
    \rho=  \sum_i \Omega_i a^{-3(1+w_i)}\, .
    \label{ecrho}
\end{equation}
This form for energy density is an ansatz  that is usually used in Eq.~\eqref{eq1} to fit the observed temporal behavior of the scale factor $a(t)$. As a result, the values of the different constants, $\Omega_i$, of the model can be obtained from the best fit to the data.

On the other hand, by using this energy density \eqref{ecrho} in the continuity equation, Eq. \eqref{eq2}, allows us to calculate the total pressure as
\begin{equation}
    p=\sum_i w_i\Omega_i a^{-3(1+w_i)}\, .
    \label{pressureG}
\end{equation}
Notice that in this simple model, $p= \sum_i p_i=\sum_i w_i\rho_i$, which recovers an extension of Dalton's law. In general, $p\not\propto \rho$, and the Universe is not filled with a simple fluid (with an ideal EoS). As shown below, the Universe exists in a more complex thermodynamic state, even when its parts are in simpler, ideal states. 

From the energy density, Eq. \eqref{ecrho}, and knowing that Eq.~\eqref{eq1} holds, the deceleration parameter for this theory can be derived as
\begin{eqnarray}
    q= \frac{1}{2}\frac{\sum_i (1+3w_i)\Omega_i a^{-1-3w_i}}{\sum_i  \Omega_i a^{-1-3w_i}}\, .
    \label{qsimple}
\end{eqnarray}
Similarly, the jerk parameter can be derived as
\begin{eqnarray}
    j= \frac{1}{2}\frac{\sum_i (1+3w_i)(2+3w_i)\Omega_i a^{-1-3w_i}}{\sum_i  \Omega_i a^{-1-3w_i}}\, .
    \label{jsimple}
\end{eqnarray}
In this way, the compressibility, $dp/d\rho$, can be obtained in general from Eq.~\eqref{GeneralgeneralEOS2} as
\begin{equation}
    \frac{dp}{d\rho} =\frac{\sum_i (1+w_i)w_i\Omega_i a^{-3w_i}}{\sum_i (1+w_i)\Omega_i a^{-3w_i}}\, .
    \label{ratio1}
\end{equation}
From here, it is clear that $dp/d\rho\neq\mbox{constant}$.
The denominator of \eqref{ratio1} is always positive, but the sign of the numerator depends on the type of contents used as a fitting model of the Universe.

For this general model, the meaning of $dp/d\rho$ can be straightforwardly understood by noting that Eqs.~\eqref{ecrho} and \eqref{pressureG} define a time-varying weighted mean \citep{visser}:
\begin{equation}
    \frac{p}{\rho}=\overline{ w(a)}=\frac{\sum_i w_i\Omega_i a^{-3(1+w_i)}}{\sum_i \Omega_i a^{-3(1+w_i)}}\, .
\end{equation}
In these terms, compressibility \eqref{ratio1} can be written as \citep{visser}
\begin{equation}
    \frac{dp}{d\rho}=\frac{\overline{ w(a)}+\overline{w(a)^2}}{1+\overline{w(a)}}\, ,
\end{equation}
implying that the time-varying variance, $\mbox{Var}(w)=\overline{w(a)^2}-\left(\overline{w(a)}\right)^2$, on the different fluid components for this model is defined as
\begin{equation}
    \mbox{Var}(w)=\left(1+\frac{p}{\rho}\right)\left(\frac{dp}{d\rho}-\frac{p}{\rho}\right)\, .
\end{equation}
As discussed, the compressibility $dp/d\rho$ measures the degree of dispersion among the different $w_i$ values and how this dispersion evolves over time.

In the following, we analyze different models in terms of their unique global EoS, by using
Eqs.~\eqref{ecrho}, \eqref{pressureG}, and \eqref{ratio1}.

\subsubsection{The flat $\Lambda$CDM model}

In this case, we have a model that proposes  four types of barotropic ideal fluids contributing to the total content of the Universe. These include nonrelativistic cold matter ($w_m=0)$, cold dark matter ($w_c=0$),
radiation ($w_R=1/3$), and dark energy produced by a cosmological constant ($w_\Lambda=-1$).
In addition, we assume that the space curvature is zero.
In such a case, the total energy density \eqref{ecrho} is written as
\begin{equation}
    \rho= \Omega_M a^{-3}+\Omega_R a^{-4}+\Omega_\Lambda\, ,
    \label{rhoflatlcdm}
\end{equation}
where $\Omega_M=\Omega_m+\Omega_c$.
Using Eq. \eqref{pressureG}, we 
can obtain the total pressure of this model as
\begin{equation}
    p=\frac{1}{3}\Omega_R a^{-4}-\Omega_\Lambda\, .
    \label{pflatlcdm}
\end{equation}
We can combine the results of Eqs. \eqref{rhoflatlcdm} and \eqref{pflatlcdm} to obtain the EoS. From the pressure derived above for this model, we find that $a=(3(p+\Omega_\Lambda)/\Omega_R)^{-1/4}$. Applying this to our definition of energy density, we obtain the EoS for the content in the $\Lambda$CDM model as
\begin{eqnarray}
\rho(p)=3p+4\Omega_\Lambda+\Omega_M\left(\frac{3}{\Omega_R}\left(p+\Omega_\Lambda\right) \right)^{3/4}\, .
\end{eqnarray}
This EoS describes a fluid that departs from the ideal gas. Thus, the flat $\Lambda$CDM model intrinsically implies that the Universe's fluid content exists in a complex thermodynamical state.
More simply, this EoS can be expressed in terms of an effective energy density $\rho_{\mathit{eff}}=\rho-\Omega_\Lambda$ and an effective pressure $p_{\mathit{eff}}=p+\Omega_\Lambda$:
\begin{equation}
\rho_{\mathit{eff}}=3p_{\mathit{eff}}+\Omega_M\left({3}p_{\mathit{eff}}/{\Omega_R} \right)^{3/4}\, .
\end{equation}
From this, we can see that the contribution of the cosmological constant corresponds to a shift in the zero values of the fluid energy density and pressure, rather than generating the anomalous pressure (negative compressibility) associated with dark energy.

This can be better understood by calculating $d\rho/dp$. Using the above EoS, we obtain
\begin{eqnarray}
    \frac{d\rho}{dp}=3+\frac{9\Omega_M}{4\Omega_R}\left(\frac{3}{\Omega_R}(p+\Omega_\Lambda) \right)^{-1/4}>0\, .
    \label{eoslambdacdm1}
\end{eqnarray}
This implies that the whole content predicted by the flat $\Lambda$CDM model is just an ordinary fluid. 

Another way to demonstrate this is by calculating the $q$ and $j$ parameters for this theory. The deceleration parameter is given by
\begin{eqnarray}
    q=\frac{\Omega_M a/2+\Omega_R-\Omega_\Lambda a^4}{\Omega_M a+\Omega_R+\Omega_\Lambda a^4}\, ,
    \label{qlambdacdm}
\end{eqnarray}
while the jerk is
\begin{eqnarray}
    j=\frac{\Omega_M a+3\Omega_R+\Omega_\Lambda a^4}{\Omega_M a+\Omega_R+\Omega_\Lambda a^4}\, .
\end{eqnarray}
Notice that the jerk parameter for the $\Lambda$CDM model is not 1. Only when the relativistic contribution is neglected
($\Omega_R=0$), does $j=1$. Therefore, using Eq.~\eqref{GeneralgeneralEOS2}, we obtain
\begin{equation}
    \frac{dp}{d\rho}=\frac{4\Omega_R}{12\Omega_R +9\Omega_M a}>0\, .
    \label{eoslambdacdm2}
    \end{equation}

 Equations~\eqref{eoslambdacdm1} and \eqref{eoslambdacdm2} are equivalent. Both indicate that the flat $\Lambda$CDM model
defines a Universe with an ordinary fluid content that initially behaves as radiation ($dp/d\rho\approx 1/3$ for $a\ll 1$) and later transitions to a nonrelativistic fluid with $dp/d\rho<1/3$ as the scale factor $a$ increases. When the relativistic contribution is neglected, then $dp/d\rho=0$, which implies that the fluid pressure is constant.
However, as $dp/d\rho > 0$,  the cosmological constant of this model introduces no anomalous phenomenon that can be associated with dark energy. 

This interpretation does not align with the usual paradigm, which associates the accelerated expansion of the Universe with the contribution of the cosmological constant to the energy density. From the perspective of a unique EoS for the fluid content of the Universe, this association is incorrect. For the $\Lambda$CDM model, the pressure of the Universe only increases if the energy density increases. To show how this conflicts with the usual paradigm, let us consider Eq.~\eqref{relationprhodivi} for $p/\rho$. In this case, using Eq. \eqref{pflatlcdm}, this ratio can be negative for this model when the scale factors are $a>a_{q1}$ via 
\begin{eqnarray}
    a_{q1}=\left(\frac{\Omega_R}{3\Omega_\Lambda}\right)^{1/4}\, .
\end{eqnarray}
For scale factors larger than this value, the pressure becomes negative in the $\Lambda$CDM model. However, this does not imply a decrease in the energy density of the Universe. It simply represents a shift in the pressure scale. 
Similarly,
it is usually said that the dominance of dark energy begins when $q<0$, producing acceleration $\ddot a>0$.  This occurs when $\rho+3p<0$, or using Eq.~\eqref{qlambdacdm}, when $2\Omega_R-2\Omega_\Lambda a^4+\Omega_M a<0$.
When $\Omega_R$ is very small compared to other cosmological quantities from Eq.~\eqref{qlambdacdm}, this occurs for scale factors, $a>a_{q2}$, where
\begin{equation}
a_{q2}=\left(\frac{\Omega_M}{2\Omega_\Lambda}\right)^{1/3}+\frac{2\Omega_R}{3\Omega_\Lambda}\, .
\end{equation} 
After this scale factor is reached, the Universe begins to show acceleration. But to achieve this, pressure must increase along with energy density.
Neither $a_{q1}$ nor $a_{q2}$ represents a transition for a system dominated by dark energy under the interpretation of the compressibility properties of the fluid content. They are just a reflection of the complexity of the fluid's EoS. 
In terms of fluid compressibility \eqref{eoslambdacdm2}, both values for dark energy dominance are incorrect and biased, as the complete dynamics is governed by  both $q$ and the jerk parameter.

In this interpretation, the accelerated expanding evolution of the Universe is due to the fact that the fluid does not exist in an ideal state. In this case, the pressure and energy density have shifted to zero scales, used to model the observed data.
In the next section, we demonstrate that this is not the case when dark energy is not a cosmological constant, but rather a dark fluid with a more complex EoS.

\subsubsection{The {$w$}CDM model}

In this fitting model, we upgraded the contribution of a cosmological constant to that of a dark energy fluid, with $-1<\omega_\Lambda<0$.
The energy density of the model becomes
\begin{equation}
    \rho=\Omega_M a^{-3}+\Omega_R a^{-4}+\Omega_\Lambda a^{-3(1+\omega_\Lambda)}+\Omega_k a^{-2}\, ,
\end{equation}
where we included the effect of spatial curvature with $\Omega_k=1-\Omega_M-\Omega_R-\Omega_\Lambda$ and  $\Omega_M=\Omega_m+\Omega_c$.
From Eq.~\eqref{pressureG}, the total pressure becomes
\begin{equation}
    p=\frac{1}{3}\Omega_R a^{-4}+\omega_\Lambda\Omega_\Lambda a^{-3(1+\omega_\Lambda)}-\frac{1}{3}\Omega_k a^{-2}\, .
\end{equation}
In this case, obtaining an explicit EoS in the form $p=p(\rho)$ is not simple. However, the relation $(\rho,p)$ can be obtained in a parametric form.
Nevertheless, calculating the derivative of this EoS from Eq.~\eqref{ratio1} is straightforward:
\begin{eqnarray}
    \frac{dp}{d\rho}=\frac{4\Omega_R+9(1+\omega_\Lambda)\omega_\Lambda\Omega_\Lambda a^{1-3\omega_\Lambda}-2\Omega_k a^2}{12\Omega_R+9\Omega_M a+9(1+\omega_\Lambda) \Omega_\Lambda a^{1-3\omega_\Lambda}+6\Omega_k a^2}\, .
    \label{wcompress}
\end{eqnarray}
Equation~\eqref{wcompress} indicates that the behavior of the $w$CDM model is more complex than that of the $\Lambda$CDM model.
In the $w$CDM model, the fluid can acquire negative compressibilities when $dp/d\rho<0$. This occurs for scale factors fulfilling 
\begin{equation}
4\Omega_R+9(1+\omega_\Lambda)\omega_\Lambda\Omega_\Lambda a^{1-3\omega_\Lambda}-2\Omega_k a^2<0\, ,
\label{condition1wcdm}
\end{equation}
as the denominator of Eq.~\eqref{wcompress} is positive.
When the curvature is neglected ($\Omega_k=0$), negative compressibility occurs for scale factors
larger than a critical value  $a>a_{crit}$,
with
\begin{equation}
    a_{crit}=\left( -\frac{4\Omega_R}{9(1+\omega_\Lambda)\omega_\Lambda\Omega_\Lambda}\right)^{1/(1-3\omega_\Lambda)}\, .
    \label{acritwcdmOk0}
\end{equation}
In this case, at early times, when $a\leq a_{crit}$, $dp/d\rho\geq0$, and the contents of the Universe behave as a normal ordinary fluid.
But later, when expansion drives $a > a_{crit}$, then $dp/d\rho < 0$, and the Universe enters a regime of negative compressibility.
Therefore, from the perspective of a fluid with a unique EoS in the $w$CDM model, we can claim that the Universe began to display dark-energy properties (in terms of an increase in pressure with decreasing energy density) from the time when $a(t) = a_{crit}$.
Notice that this point in time depends only on $\Omega_\Lambda$ and $\Omega_R$ for a flat Universe and exists because $-1\neq\omega_\Lambda<0$.
Furthermore, the $w$CDM model predicts that once the fluid content reaches a phase of negative compressibility, the Universe will never return a normal fluid state with positive compressibility.  

On the other hand, if the radiation epoch is neglected ($\Omega_R=0$), then negative compressibility is achieved at all times for positive spatial curvatures ($\Omega_k>0$). In the opposite case, for negative curvatures ($\Omega_k<0$), negative compressibility appears for scale factors larger than the critical value
\begin{equation}
    a_{crit}=\left(\frac{9(1+\omega_\Lambda)\omega_\Lambda\Omega_\Lambda}{2\Omega_k} \right)^{1/(3\omega_\Lambda-1)}.
\end{equation}

By contrast, we can compare this critical scale factor with that obtained from the condition, $p/\rho<0$ (or $q<1/2$). This is fulfilled
for scale factors 
satisfying the condition
\begin{eqnarray}
    \Omega_R+3\omega_\Lambda\Omega_\Lambda a^{1-3\omega_\Lambda}-\Omega_k a^2<0\, .
    \label{condition2wcdm}
\end{eqnarray}
In the case of negligible spatial curvature, any scale factor
$a>a_{q1}$ greater than
\begin{eqnarray}
    a_{q1}=\left(-\frac{\Omega_R}{3\omega_\Lambda\Omega_\Lambda}\right)^{1/(1-3w_\Lambda)}\, 
    \label{scaleprhonegativa}
\end{eqnarray}
produces $p/\rho<0$.
Similarly,
this critical scale factor can contrast
with the value derived from the condition $q<0$
for the deceleration parameter. This condition translates to
\begin{eqnarray}
    \Omega_M a+2\Omega_R+(1+3\omega_\Lambda)\Omega_\Lambda a^{1-3\omega_\Lambda}<0\, .
    \label{condition3wcdm}
\end{eqnarray}
For the case of flat curvature and very small $\Omega_R$, $q<0$ implies a scale factor, 
\begin{eqnarray}
    a_{q2}=\left(-\frac{\Omega_M}{\Omega_\Lambda(1+3\omega_\Lambda)}\right)^{-1/(3w_\Lambda)}- \frac{2\Omega_R}{3\omega_\Lambda\Omega_\Lambda}\, ,
    \label{deacellwcdmneative}
\end{eqnarray}
such that for $a>a_{q2}$ the deceleration parameter becomes negative.

From the point of view of a unique EoS, dark energy dominates when compressibility becomes negative. This occurs neither when $p/\rho$ is negative nor when deceleration is negative. As $\Omega_R$ is very small, then 
$a_{crit}$ could be larger or smaller than $a_{q1}$ or $a_{q2}$, depending on the value of $\omega_\Lambda$. However, it is very unlikely that these three scale factors coincide.  

 In general, a nonzero value for $\Omega_R$ is a critical condition for negative compressibility for flat models. Otherwise, $a_{q2}$ remains the sole marker for inferring the existence of dark energy. Since we know that the jerk parameter provides information on the evolution of dark energy, we need to consider $\Omega_R\neq 0$ in the above models.
For instance, assuming the main values, $\Omega_M= 0.298$ and $\omega_\Lambda= -0.912$, given in \citet{Abdul} and considering $\Omega_R\approx 8.25\times 10^{-5}$ \citep{carrol} for a flat model ($\Omega_k=0$), then $\Omega_\Lambda\approx 0.702$. With these values, the critical scale factor from Eq. \eqref{acritwcdmOk0},
where the fluid acquires negative compressibility, is $a_{crit}\approx 0.14$. On the other hand, the scale factor from Eq. \eqref{scaleprhonegativa}, where $p/\rho<0$ is $a_{q1}\approx 0.07$. Moreover, the scale factor from Eq. \eqref{deacellwcdmneative}, where $q$ becomes negative, is $a_{q2}\approx 0.6$. As $a_{crit}<a_{q2}$ in the $w$CDM model when $\Omega_R$ is taken into account, the dark energy effect (in terms of negative compressibility) occurs earlier than predicted by the negative value of the deceleration parameter.

\subsection{Examples of sophisticated  models}

A similar analysis can be performed for more complex fitting models for energy density. Again, for any general form of this model, the pressure may be  
obtained from Eq.~\eqref{equooressureGenrrr}, while compressibility may be obtained from Eq.~\eqref{GeneralgeneralEOS2}.

\subsubsection{The $w_0$$w_a$CDM model}

Let us define the energy density of the content of the Universe as
\begin{equation}
    \rho=\Omega_M a^{-3}+\Omega_R a^{-4}+\Omega_\Lambda\,  I(a)\, ,
\end{equation}
where $I$ is an arbitrary function of the scale factor. In this case, the pressure derived from Eq. \eqref{equooressureGenrrr} is
\begin{eqnarray}
    p=\frac{1}{3}\Omega_R a^{-4}-\Omega_\Lambda\left(\frac{a}{3}I'+I\right)\, ,
    \label{pressurewowacdm}
\end{eqnarray}
where   $I'\equiv dI/da$. Also,
the deceleration parameter is
\begin{eqnarray}
    q=\frac{\Omega_M a/2+\Omega_R-\Omega_\Lambda a^3 Y'/2}{\Omega_M a+\Omega_R+\Omega_\Lambda a^2 Y}\, ,
\end{eqnarray}
where $Y=a^2 I$. Similarly, the jerk parameter becomes
\begin{eqnarray}
    j=\frac{\Omega_M a+3\Omega_R+\Omega_\Lambda a^4 Y''/2}{\Omega_M a+\Omega_R+\Omega_\Lambda a^2 Y}\, ,
\end{eqnarray}
where  $Y''\equiv d^2Y/da^2$. Thus, the compressibility \eqref{GeneralgeneralEOS2}, in terms of $I$, becomes 
\begin{eqnarray}
    \frac{dp}{d\rho}=\frac{4\Omega_R+3\Omega_\Lambda a^5\left( aI''+4I' \right)}{9\Omega_Ma+12\Omega_R-3\Omega_\Lambda a^5 I'}\, .
    \label{comprewowacdm}
\end{eqnarray}

The compressibility of this fluid content depends on the exact form of $I(a)$.
The $w_0$$w_a$CDM model can be obtained when the dark energy fluid component is assumed to have a variable EoS, with a linear dependence: $w_\Lambda(a)=w_0-w_a-w_a a$, where $w_0$ and $w_a$
are constant ($-1<w_0<0$). This is equivalent to $w_\Lambda(z)=w_0+w_a z/(1+z)$, in terms of redshift $z=1/a-1$. Thus, 
for the $w_0$$w_a$CDM model,
 we find \citep{shi}
\begin{eqnarray}
    I(a)=a^{-3(1+w_0+w_a)}e^{-3w_a(1-a)}\, .
\end{eqnarray}
Using this expression, the pressure, Eq.~ \eqref{pressurewowacdm}, in the $w_0$$w_a$CDM model can be succinctly written as
\begin{equation}
    p=\frac{1}{3}\Omega_R a^{-4}-\Omega_\Lambda\left(w_0+w_a(1-a) \right)I\, ,
    \label{pressurewowacdm2}
\end{equation}
whereas its compressibility \eqref{comprewowacdm} is
\begin{eqnarray}
    \frac{dp}{d\rho}=\frac{4\Omega_R+3\Omega_\Lambda a^5I+9a^4\Omega_\Lambda I\left(w_0+w_a(1-a)\right)\left(1+w_0+w_a(1-a) \right)}{9\Omega_Ma+12\Omega_R+9\Omega_\Lambda a^4 I(1+w_0+w_a(1-a))}\, .
    \label{comprewowacdm2}
\end{eqnarray}

We can analyze  the acceleration  condition, $q<0$ ($\ddot a>0$), for the $w_0$$w_a$CDM model. This is achieved when $\rho+3p<0$, which translates into
\begin{eqnarray}
   2\Omega_R+ \Omega_M a+\Omega_\Lambda a^4 I\left(1+3w_0+3w_a(1-a)\right)<0 \, .
 \label{comprewowacdm4}
 \end{eqnarray}
This condition can be satisfied for $-1<w_0<0$ and, for instance, $w_a>0$.
It defines the scale factor in which the Universe begins its accelerated expansion. However, this is not the same scale factor that produces negative compressibility of the content fluid. In the case of $-1<w_0<0$ and $w_a>0$, compressibility, expressed via Eq. \eqref{comprewowacdm2}, becomes negative for the scale factor satisfying
\begin{eqnarray}
   && 4\Omega_R+3\Omega_\Lambda a^5I\nonumber\\
    &&+9a^4\Omega_\Lambda I\left(w_0+w_a(1-a)\right)\left(1+w_0+w_a(1-a) \right)<0\, .
    \label{comprewowacdm3}
\end{eqnarray}
The conditions in Eqs. \eqref{comprewowacdm4} and  \eqref{comprewowacdm3} are quite different. They define two scale factors with distinct meanings. Equation~\eqref{comprewowacdm4} predicts when the Universe accelerates, but not its cause. This is measured by the negative compressibility condition (Eq. \eqref{comprewowacdm3}), which provides the corresponding time at which dark energy produces the necessary changes in the Universe's content to produce acceleration.

\subsubsection{The generalized Chaplygin gas model}
 
In this model, the energy density of the Universe's content is given by \citep{Panotopoulos,Kamenshchik,shi,zzhu}
 \begin{eqnarray}
     \rho= \Omega_b a^{-3}+\Omega_\Lambda\left[(1+w)a^{-3(1+\alpha)}-w\right]^{1/(1+\alpha)}\, ,
     \label{densityChaplygin}
 \end{eqnarray}
where
$\Omega_b+\Omega_\Lambda=1$, $-1<w<0$, and $\alpha>0$. This model is constructed assuming one of the Universe's components is an exotic fluid with EoS $p\propto -\rho^{-\alpha}$ \citep{Panotopoulos,Kamenshchik,shi,zzhu}. From Einstein Eqs.~\eqref{eq2} and \eqref{equooressureGenrrr}, the total pressure of this gas system is given by
\begin{eqnarray}
     p= \omega\, \Omega_\Lambda\left[(1+w)a^{-3(1+\alpha)}-w\right]^{-\alpha/(1+\alpha)}<0\, .
     \label{pressureChaplygin}
 \end{eqnarray}
This suggests that the Universe has a fluid content with negative pressure across its evolution. The EoS of this gas can be readily obtained as
\begin{eqnarray}
    \rho=\Omega_b\left( \frac{w+(p/\Omega_\Lambda\omega)^{-(1+\alpha)/\alpha}}{1+w}\right)^{-1/(1+\alpha)}+\Omega_\Lambda\left(\frac{p}{\Omega_\Lambda\omega}\right)^{-1/\alpha}\, ,
\label{eoschaplygin}
\end{eqnarray}
which indicates its nonideal nature.

Furthermore, we can calculate the deceleration and jerk parameters of this theory and use them in Eq.~\eqref{GeneralgeneralEOS2} to obtain the compressibility of the Chaplygin gas:
\begin{equation}
    \frac{dp}{d\rho}=\frac{-\alpha w(1+w)\Omega_\Lambda a^{-3\alpha}\left[(1+w)a^{-3(1+\alpha)}-w\right]^{-(1+2\alpha)/(1+\alpha)}}{\Omega_b+\Omega_\Lambda (1+w) a^{-3\alpha}\left[(1+w)a^{-3(1+\alpha)}-w\right]^{-\alpha/(1+\alpha)}}>0\, .
\end{equation}
The compressibility of the whole Universe described by the generalized Chaplygin gas is always positive. 
Therefore, similar to that which occurs in the $\Lambda$CDM model, this fluid does not present any anomalous behavior that can be characterized with a dark energy epoch. This occurs even under negative pressure, i.e. Eq. \eqref{pressureChaplygin}, which represents a type of pressure scale rather than a description of fluid behavior. The accelerated behavior of the model's scale factor results from the EoS \eqref{eoschaplygin}, which describes a complicated -- rather than ideal --fluid.
 In fact, we can calculate when the Universe accelerates using $q<0$ (or
$\rho+3 p<0$) from Eqs. \eqref{densityChaplygin} and \eqref{pressureChaplygin}. This condition is expressed as
\begin{equation}
   \frac{\Omega_b}{\Omega_\Lambda} a^{-3}+\left[(1+w)a^{-3(1+\alpha)}-w\right]^{-\alpha/(1+\alpha)}\left[(1+w)a^{-3(1+\alpha)}+2w\right]<0\, , 
\end{equation}
which can be fulfilled for $-1<w<0$. Thus, at a given scale factor, this Universe exhibits acceleration $\ddot a>0$, as  both pressure and energy increase.

\subsubsection{The DGP model with a self-accelerating branch}

Lastly, we describe the fluid EoS in the Dvali-Gabadadze-Porrati (DGP) model \citep{TMDavis, ALue,Panotopoulos}. In this case, the fluid energy density is given by
\begin{eqnarray}
\rho=\left( \sqrt{\Omega_R}+\sqrt{\Omega_M a^{-3}+\Omega_c}\right)^2\, ,
    \label{densitydgp}
\end{eqnarray}
where $\Omega_c=(1-\Omega_M)^2/4$. 
Using Eq.~\eqref{equooressureGenrrr} we obtain the total fluid pressure in the modeled Universe as
\begin{eqnarray}
    p=-\sqrt{\Omega_R}\left(\frac{\Omega_M a^{-3}+2\Omega_c}{\sqrt{\Omega_M a^{-3}+\Omega_c}}\right)-\Omega_R-\Omega_c <0\, .
    \label{pdensitydgp}
\end{eqnarray}
Thus, the fluid pressure defined in the DGP model is negative. We can combine the two equations above to obtain the EoS of this fluid via
\begin{eqnarray}
    p=-\sqrt{\rho}\left(\sqrt{\Omega_R}+\frac{\Omega_c}{\sqrt{\rho}-\sqrt{\Omega_R}}\right)<0\, .
\end{eqnarray}
This expression describes the nonideal feature of this fluid and its pressure negativity (as $\sqrt{\rho}>\sqrt{\Omega_R}$). Subsequently, we can determine compressibility \eqref{GeneralgeneralEOS2} by calculating the $q$ and $j$ parameters from this model via
\begin{eqnarray}
    \frac{dp}{d\rho}=-\frac{\sqrt{\Omega_R}\, \Omega_M a^{-3}}{2\sqrt{\rho}(\Omega_M a^{-3}+\Omega_c)}<0\, .
\end{eqnarray}
Accordingly, the fluid consistently displays an increase in pressure with a decrease in energy density.
Therefore, even without an explicit dark energy component, the DPG Universe consistently behaves as an anomalous fluid with negative compressibility as if it were dominated by dark energy across its  whole evolution. 

 To demonstrate the accelerated  behavior of the scale factor of this Universe, we can calculate the condition, $\rho+3p<0$, from Eqs. \eqref{densitydgp} and \eqref{pdensitydgp}. This condition reads
\begin{equation}
    \Omega_M a^{-3}<2\Omega_R+2\Omega_c+\frac{\sqrt{\Omega_R}\left(\Omega_M a^{-3}+4\Omega_c\right)}{\sqrt{\Omega_M a^{-3}+\Omega_c}}\, ,
\end{equation}
for a given scale factor. This Universe experiences an accelerating transition in time even when pressure is always decreasing as energy density is increasing.

\section{Discussions}
\label{se:discussion}

This work is devoted to characterizing the dark energy properties of a Universe through the EoS of its fluid content rather than  using the temporal evolution of its scale factor. This analysis differs from the usual understanding of dark energy. It adds no new physics nor new models but rather uses the EoS defined by each model -- an under-explored avenue in dark energy studies.

From our calculations above, we highlight three main points. First, the features assigned to dark energy, which cause space-time pressure expansion as a fluid's energy density decreases, can arise directly from a fluid's EoS. Therefore, as demonstrated in other fields of physics (such as condensed matter), we can characterize this anomalous behavior through the sign of compressibility. If this case holds, then several models used to account for dark energy in observational data are biased, since they do not show the anomalous behavior of compressibility in their modeled fluid content. In fact, they behave as ordinary fluids. The accelerated behavior of space-time, which results from the fluid EoS used in these models, is clearly  a nonideal effect.

Second, the complete accelerated behavior of space-time cannot be determined solely by the deceleration parameter $q$. The jerk parameter $j$
plays a fundamental role, as it is essential for measuring compressibility. Any estimation using only the $q$ parameter may produce spurious results. Furthermore, according to the information gathered on fluid properties, no additional higher-order parameter (in the scale factor expansion) is needed to understand the normal or anomalous behavior induced in cosmological space-time.

Third, compressibility measures the dispersion on the dataset used to calculate the deceleration parameter. From a statistical point of view,  compressibility is essential to obtain its variance.
It contains information that cannot be neglected and intrinsically differs from that in the simpler relation, $p/\rho$.

With this in mind, a different understanding of current dark energy models must be considered. For instance, although $\Lambda$CDM and Chaplygin gas  models are used to fit the dark energy behavior of observational data, we demonstrated that these models define the Universe's content without anomalous (dark energy) properties. Nevertheless, the proposed complex, nonideal -- but ordinary --
fluid content produces the accelerated scale factor. By contrast, the DGP model, which defines new forms of dark energy coupling,  uses a fluid that is always anomalous. Thus, it is straightforward that this model can fit dark energy.
Finally, more complicated models, such as $w$CDM or $w_0 w_a$CDM, propose specific transitions for the scale factor, in which an ordinary fluid becomes anomalous, thereby triggering the effect of dark energy (with an increase in fluid pressure and a decrease in energy density). Importantly, these transitions depend on the contribution from the radiation epoch, which cannot be ignored. Due to this contribution, the models predict that the anomalous behavior of the cosmological fluid, and therefore its effect on the scale factor, will dominate sooner than expected. 
These new predictions can be compared with future measurements to exclude some models.

Lastly, it is important to note that the formalism for compressibility \eqref{GeneralgeneralEOS2} can be 
used to study several fluid properties in any cosmological model. Following our outlined procedure, other properties may be studied by using higher-order derivatives of the scale factor. This includes  cosmological fluid phase transitions predicted by different models, which can be evaluated through variations in the speed of sound, $d^2p/d\rho^2$. In this case, for a given cosmological model, ${d^2p}/{d\rho^2}=(1/\dot a)\left({d\rho}/{da}\right)^{-1}\left({d^2p}/{dtd\rho}\right)$.
Using Eq.~\eqref{eq1}, we find the explicit relation,
 \begin{equation}
    \frac{d^2p}{d\rho^2}=-\frac{4\pi G}{9H^2(1+q)^3}\left[s(1+q)+j(1+j+4q+q^2)+q(1+2q) \right]\, , 
 \end{equation}
which introduces the snap parameter, $s=a^3\, \ddddot{a}/\dot a^4$. This expression was found in~\citet{visser} in an approximated context; however, we note that it is indeed exact.
To evaluate whether a cosmological model shows a phase transition through its fluid content ansatz, both conditions, $dp/d\rho=0$
and $d^2p/d\rho^2=0$, must be fulfilled. This implies that cosmological phase transitions occur only at specific times when they are simultaneously
fulfilled:
\begin{eqnarray}
    j=1\, ,\qquad \mbox{and}\, \qquad s+3q=-2\, ,
\end{eqnarray}
such that $q\neq -1$.
Thus, these conditions must be satisfied  at some specific  scale factor. For this, the snap parameter must be determined in order to study the fluid characteristics related to a cosmological phase transition. In particular, the flat $\Lambda$CDM and flat $w$CDM models do not fulfill these conditions and thereby do not present cosmological phase transitions.
We leave the analysis of fluid phase transitions, among other properties, for general cosmological models to future work.

Finally, we conclude that in any cosmological model, the content of the Universe  has a single EoS that departs from the one of an ideal fluid. It is from this EoS where space-time inherits its complex properties. This contrasts with the standard view of cosmology, in which content is modeled as a collection of several ideal gases, with the global characteristics of the Universe assumed to arise from only one  of them. A correct interpretation will be clarified by future observations capable of determining the jerk parameter with sufficient precision.

\begin{acknowledgements}
FAA thanks Paula Mellado for useful discussions and to FONDECYT grant No. 1230094 that partially supported this work.
OR and AC acknowledge partial funding from ANID/FONDECYT grant 1251692.
OR is supported by the Rubin-Chile Fund under grant DIA2650.
 \end{acknowledgements}

\end{document}